\documentclass[12pt]{article}
\usepackage{amsfonts}
\newcommand{\Nl}{\mathbb{N}}
\newcommand{\Ir}{\mathbb{Z}}

\newtheorem{theorem}{Theorem}[section]
          
\newtheorem{lemma}[theorem]{Lemma}

\newtheorem{definition}[theorem]{Definition}

\newcommand{\Section}[1]{\setcounter{equation}{0}\section{#1}}

\newenvironment{proof}{\noindent {\bf Proof: }}{\QED\medskip}
\def\QED{{\hspace*{\fill}{\vrule height .5ex width 1ex }\quad} 
    \vskip 0pt plus20pt}

\newcommand{\be}{\begin{equation}}
\newcommand{\ee}{\end{equation}}
\newcommand{\bea}{\begin{eqnarray}}
\newcommand{\eea}{\end{eqnarray}}
\newcommand{\beann}{\begin{eqnarray*}}
\newcommand{\eeann}{\end{eqnarray*}}
\DeclareMathAlphabet{\mathol}{OT1}{cmr}{l}{ol}

\begin{document}

\begin{center}
{\LARGE \bf A General Method for Obtaining a Lower Bound for the Ground State Entropy Density of the Ising Model With Short Range Interactions\\[27pt]}
{\large \bf Lawrence Pack and Ram Anand Puri\\[10pt]}
{\large  Department of Mathematics\\
University of California, Davis\\
One Shields Avenue\\
Davis, CA 95616-8633, USA\\[15pt]}
{\normalsize lwpack@math.ucdavis.edu}\\
{\normalsize puri@math.ucdavis.edu}\\
\vspace{.5cm}
\end{center}

\noindent
{\bf Abstract:}  We present a general method for obtaining a lower bound for the ground state entropy density of the Ising Model with nearest neighbor interactions.  Then, using this method, and with a random coupling constant configuration, we obtain a lower bound for the ground state entropy density of the square, triangular, and hexagonal two-dimensional lattices with free, cylindrical, and toroidal boundary conditions.

\Section{Introduction}

\subsection{Overview}

In this paper, we establish a general method for obtaining a lower bound for the ground state entropy density $\hat{S_0}$ of arbitrary lattices with arbitrary boundary conditions.  We then use this method on two-dimensional square, triangular, and hexagonal lattices with free, cylindrical, and toroidal boundary conditions---while assuming a random coupling constant configuration---to obtain numerical values for these lower bounds.  Let us now note that our method to be presented is derived from a method employed by Avron \emph{et al} [1], in which a value of approximately $1.100 \times 10^{-19}$ is obtained as a lower bound for $\hat{S_0}$ for the two-dimensional square lattice with a random coupling constant configuration.  Using our method, we obtain a value of approximately $4.883 \times 10^{-6}$ for this same lattice.  Let us now also note that Loebl and Vondrak [2] obtained an upper bound of $0.548$ for $\hat{S_0}$ for a certain class of square lattices with toroidal boundary conditions.  Loebl and Vondrak then conjecture that this upper bound holds for all square lattices with toroidal boundary conditions.  If this is indeed true, then coupled with our result, we find that the ground state entropy density of the square lattice with toroidal boundary conditions and a random coupling constant configuration is bounded by \[4.883 \times 10^{-6} \leq \hat{S_0} \leq 0.548.\]  However, it should be noted that while our lower bound of $4.883 \times 10^{-6}$ is many orders of magnitude greater than the previous lower bound of $1.100 \times 10^{-19}$, it is still far from what one would expect based on the upper bound of $0.548$.  This discrepancy is most likely attributable to the method which we use in this paper to establish the lower bound.

\subsection{Background Information}

The Ising Model is defined by a finite set $\Lambda$, representing atoms, and a \emph{coupling constant configuration} $J: B(\Lambda) \rightarrow \{-1,1\}$, describing the interaction between neighboring atoms.  The set $B(\Lambda)$ is the set of all \emph{bonds} in $\Lambda$.  Formally, we write $B(\Lambda) \subseteq \{ \{s_i, s_j\} : s_i,s_j \in \Lambda, i \neq j \}$.  We call elements of $\Lambda$ \emph{sites}, and often denote the bond between the sites $s_i,s_j$ by $b_{ij}$.  Note that with this definition of the Ising Model, we are not restricting ourselves to only subsets of $\Ir^2$, but in fact are considering arbitrary graphs.  A \emph{spin state} for $\Lambda$ is an assignment of either spin up or spin down to each atom; that is, a spin state is a function $\sigma: \Lambda \rightarrow \{-1,1\}$.  The energy of the system is given by $$H(\Lambda,J,\sigma) = -\sum_{b_{ij} \in B(\Lambda)} J_{ij} \sigma_i \sigma_j.$$  The \emph{ground state energy} is the minimum possible energy.  Let us now note that if $J_{ij} \sigma_i \sigma_j = -1$ for some bond $b_{ij}$, then this in fact increases the energy; hence, we call such a bond \emph{unhappy}.  The set of all unhappy bonds in $\Lambda$ with $J$ and $\sigma$ is denoted $U(\Lambda,J,\sigma)$.  Given $\Lambda$ with a $J$, a spin state $\sigma_0$ that yields the ground state energy is called a \emph{ground state}.   The set of all ground states is denoted by $D_0(\Lambda,J)$, and the number of ground states is called the \emph{ground state degeneracy}.  We will henceforth be considering only connected sets $\Lambda$ with a given $J$.   However, to properly define connectedness, we must first introduce the notion of a curve.

\begin{definition}
Given $\Lambda$, a \emph{curve of length} $n \geq 1$ in $\Lambda$ is a finite sequence $s: \{0,1, \ldots ,n\} \rightarrow S$, where $S \subseteq \Lambda$ and $\{s_{i-1},s_i\} \in B(\Lambda)$ for all $1 \leq i \leq n$.  The curve $s$ is said to {\em connect} the sites $s_0$ and $s_n$, and the set of all curves connecting $s_0$ and $s_n$ is denoted by $\mathcal{C} (s_0,s_n)$.  Finally, we call $s$ a \emph{closed curve} iff $s_n = s_0$.
\end{definition}

The notion of a closed curve will become useful in the next section.  We can now say what me mean by connectedness.  By a set $\Lambda$ being \emph{connected}, we mean that for any two sites $x,y \in \Lambda$, there exists a curve $c \in \mathcal{C} (x,y)$ connecting them.  

In this paper, we obtain a lower bound for the ground state entropy density; to do this, however, we must first obtain a lower bound for the ground state degeneracy, which in turn, requires the following three preliminary results to establish our formalism.

\Section{Preliminary Results}

Our first result is the following trivial theorem which tells us that in order to minimize the energy we need to minimize the number of unhappy bonds.

\begin{theorem}
\label{degen}
Given a spin state $\sigma$ for $\Lambda$,  $H(\Lambda,J,\sigma) = 2|U(\Lambda,J,\sigma)| - |B(\Lambda)|$.
\end{theorem}

\begin{proof}
Letting $U = U(\Lambda,J,\sigma)$, we have
\begin{eqnarray*} H(\Lambda,J,\sigma) & = & -\sum\limits_{b_{ij} \in B(\Lambda)}J_{ij}\sigma_i\sigma_j = -\left (\sum\limits_{b_{ij} \in B(\Lambda) \cap U}J_{ij}\sigma_i\sigma_j + \sum\limits_{b_{ij} \in B(\Lambda) \setminus U}J_{ij}\sigma_i\sigma_j \right ) \\ & = & -\left (\sum\limits_{b_{ij} \in B(\Lambda) \cap U} -1 + \sum\limits_{b_{ij} \in B(\Lambda) \setminus U} 1 \right ) \\ & = & -(-|U(\Lambda,J,\sigma)| + |B(\Lambda)| - |U(\Lambda,J,\sigma)|) \\ & = & 2|U(\Lambda,J,\sigma)| - |B(\Lambda)|.
\end{eqnarray*}
\end{proof}

The next observation is a simple one, but fundamental to understanding our formalism.

\begin{theorem} 
\label{basis} 
Given a spin state $\sigma$ for $\Lambda$.  Then, for any closed curve $s$ in $\Lambda$ of any length $n$, we have
\[ \prod \limits_{i = 0}^{n - 1} J_{i,i + 1}\sigma_i\sigma_{i + 1} = \prod \limits_{i = 0}^{n - 1} J_{i,i + 1} .\]
\end{theorem}

\begin{proof}
This is true since
\begin{eqnarray*} \prod \limits_{i = 0}^{n - 1} J_{i,i + 1}\sigma_i\sigma_{i + 1} & = & (J_{01}\sigma_0\sigma_1)(J_{12}\sigma_1\sigma_2)\cdot\ldots\cdot(J_{n - 2,n - 1}\sigma_{n - 2}\sigma_{n - 1})(J_{n - 1,n}\sigma_{n - 1}\sigma_n) \\
 & = & \left (\prod \limits_{i = 0}^{n - 1} J_{i,i + 1} \right )\left (\prod \limits_{i = 0}^{n - 1} \sigma_i^2 \right ) \\
  & = & \prod \limits_{i = 0}^{n - 1} J_{i,i + 1}.
\end{eqnarray*}
\end{proof}

Theorem \ref{basis} tells us how the distribution of unhappy bonds depends on $J$: along each closed curve in $\Lambda$, the parity of the number of negative $J$ bonds equals the parity of the number of unhappy bonds.  This observation will be used extensively in our method of bounding the ground state degeneracy.

We are now ready for our third preliminary observation which treats the following general question.  Given a spin state $\sigma$ for $\Lambda$, what are all the spin states that give the lattice the same energy as $\sigma$?

To begin, let $S$ be any subset of $\Lambda$, and let $\sigma_S$ be the spin state for $\Lambda$ obtained from $\sigma$ by flipping the spins of the sites in $S$.  How do the lattice energies $H(\Lambda,J,\sigma)$ and $H(\Lambda,J,\sigma_S)$ compare?

This question can be answered by defining the set $B_S = \{b_{ij} \in B(\Lambda): s_i \in S, s_j \notin S \}$.  The set $B_S$ can be easily visualized by a \emph{contour}, as in Picture 1, where we have only shown the special case that $\Lambda$ is a subset of a square lattice.  Note that by flipping the spins of only the sites in $S$, only the happiness of bonds in the set $B_S$ change.  Indeed, for if the spins of $s_i$ and $s_j$ are both flipped, the happiness remains unchanged.  Moreover, if neither the spin of $s_i$ nor $s_j$ is flipped, then the happiness trivially remains unchanged.  Thus, letting $U = U(\Lambda,J,\sigma)$, to get the set $U_S = U(\Lambda,J,\sigma_S)$ of unhappy bonds in $\Lambda$ with $J$ and $\sigma_S$, we simply ``complement'' the bonds along the contour; that is,
\[U_S = [U \setminus (B_S \cap U)] \cup (B_S \setminus U).\]
From this, it immediately follows that $|U_S| = |U| - |B_S \cap U| + |B_S \setminus U|$, which coupled with Theorem \ref{degen} tells us how the lattice energies compare.  We have thus proved the following theorem.

\begin{theorem}
\label{entropic set theorem}
Given $\Lambda$ with $J$ and $\sigma$, let $U = U(\Lambda,J,\sigma)$.  For any $S \subseteq \Lambda$, define $\sigma_S$ and $U_S$ as above.

Then, $H(\Lambda,J,\sigma) = H(\Lambda,J,\sigma_S)$ iff $|B_S \cap U| = |B_S \setminus U|$.
\end{theorem}

\setlength{\unitlength}{0.5cm}
\begin{figure}[ht]
\begin{picture}(10,15.5)(-11,-0.5)


\put(2.5,14){\large Picture 1}
\multiput(0,4)(1,0){10}{\line(0,1){9}}
\multiput(0,4)(0,1){10}{\line(1,0){9}}
\multiput(-2,9)(0,1){3}{\line(1,0) {2}}
\multiput(-2,9)(1,0){2}{\line(0,1) {2}}
\put(-1,6){\line(0,1) {3}}
\multiput(-1,6)(0,1){3}{\line(1,0){1}}
\multiput(9,7)(0,1){2}{\line(1,0) {2}}
\multiput(9,7)(1,0){3}{\line(0,1) {1}}
\put(10,10){\line(0,1){2}}
\multiput(9,10)(0,1){3}{\line(1,0) {1}}
\multiput(4,11)(1,0){4}{\circle* {0.4}}
\multiput(5,12)(1,0){2}{\circle* {0.4}}
\put(6,10){\circle*{0.4}}
\put(3,6){\circle*{0.4}}

\linethickness{1.0mm}

\put(3.5,10.5){\line(0,1){1}}
\put(3.5,11.5){\line(1,0){1}}
\put(4.5,11.5){\line(0,1){1}}
\put(4.5,12.5){\line(1,0){2}}
\put(6.5,11.5){\line(0,1){1}}
\put(6.5,11.5){\line(1,0){1}}
\put(7.5,10.5){\line(0,1){1}}
\put(6.5,10.5){\line(1,0){1}}
\put(6.5,9.5){\line(0,1){1}}
\put(5.5,9.5){\line(1,0){1}}
\put(5.5,9.5){\line(0,1){1}}
\put(3.5,10.5){\line(1,0){2}}
\multiput(2.5,5.5)(0,1){2}{\line(1,0){1}}
\multiput(2.5,5.5)(1,0){2}{\line(0,1){1}}

\put(-4.7,1.5){\parbox{10cm}{The contour is shown by the thick line.  The sites of $S$ are those enclosed by the contour.  The set $B_S$ is the set of bonds that cut the contour perpendicularly.}}

\end{picture}
\end{figure}

Thus, to find the set of all spin states that give the lattice the same energy as $\sigma$, we need to find all subsets $S$ of $\Lambda$ such that the number of bonds in $B_S$ that are in $U$ is equal to the number of bonds in $B_S$ that are not in $U$; that is, $|B_S \cap U| = |B_S \setminus U|$.  We call such a set $S$ that satisfies $|B_S \cap U| = |B_S \setminus U|$ \emph{entropic} relative to $U$, or entropic relative to $\sigma$.
  
\Section{Main Results}

\subsection{Local Degeneracy}

We can now discuss the idea behind the paper.  We construct a $J$ configuration in a subset $M$ of $\Lambda$ that guarantees a local degeneracy.  By this we mean that we are guaranteed to find degeneracy within this subset independent of the $J$ configuration outside of this subset.  We call such an entity a \emph{module}.  We then suppose to have a random $J$ configuration for $\Lambda$, and then using probabilistic methods we determine the number of expected modules in all of $\Lambda$, thereby obtaining a lower bound for the ground state degeneracy.  

We now proceed to rigorously define what a module is.

\begin{definition}
A \emph{module in $\Lambda$} is a set $M \subseteq \Lambda$ with a $J_M$ such that for any $J$ for $\Lambda$ and any ground state $\sigma \in D_0(\Lambda,J)$, if $J_M = J|_{B(M)}$, then there exists a nonempty set $S \subseteq M$ such that $\sigma_S \in D_0(\Lambda,J)$.
\end{definition}

That is, we take a set of sites $M$ in $\Lambda$ and define what the coupling constant configuration for it is.  We then require $M$ to have the property that given any ground state $\sigma$, there exists at least one non-trivial entropic set relative to $\sigma$ \emph{contained in} $M$.  From this it immediately follows that if we have a collection of $n$ pairwise disjoint modules in $\Lambda$, then a lower bound for the ground state degeneracy for $\Lambda$ with any $J$ is $|D_0(\Lambda,J)| \geq 2^n$.

Let us now suppose the existence of a module $M$.  We will now demonstrate the method by which we obtain a lower bound for the ground state degeneracy.  We first let $k$ be the maximum number of pairwise-disjoint subsets $T_i$ of $\Lambda$, where each $T_i$ has the possibility of being the module $M$.  Note that if $\{T_i\}_{i = 1}^k$ is a partition of $\Lambda$, then $k = \frac{|\Lambda|}{|M|}$.  We now let $p$ be the probability that an arbitrary bond is a negative $J$ bond.  Then, the probability of any $T_i$ being the module $M$ will be some function of $p$ that depends on $M$.  We will denote this probability by $f(p)$.  We also want to consider site and bond diluted lattices.  To understand diluted lattices, it is helpful to begin with a ``first'' lattice $\Lambda_1$ with the set of bonds $B(\Lambda_1)$, and two probability functions $P_s$ and $P_b$---$P_s$ defined on $\Lambda_1$ and $P_b$ defined on $B(\Lambda_1)$.  With this set up, a \emph{site diluted lattice} is any subset $\Lambda$ of $\Lambda_1$ that is chosen consistent with $P_s$, where the interpretation of $P_s$ is that $P_s(x) = P[x \in \Lambda]$.  A site diluted lattice can be used to model a substance that has sporadic impurities, but that is homogeneous otherwise.  Semi-conducting silicon wafers doped with boron is one example.  Next, $\Lambda$ is a \emph{bond diluted lattice} if its set of bonds $B(\Lambda) \subseteq B(\Lambda_1)$ is chosen consistent with $P_b$, where the interpretation of $P_b$ is that $P_b(x) = P[x \in B(\Lambda)]$.  A bond diluted lattice can be used to model an imperfect crystal; here, there are some sets $\{s,s^{\prime}\}$ of neighboring atoms that interact in the perfect crystal $\Lambda_1$, but that do not interact in the imperfect crystal $\Lambda$, which is in fact what makes $\Lambda$ imperfect.  We will suppose that $P_s$ and $P_b$ are constant functions, and denote their values by $p_s$ and $p_b$, respectively.  With these definitions, it now follows that the probability that an arbitrary $T_i$ is the module $M$ is given by $f(p) p_s^{|M|} p_b^{|B(M)|}$.  With this in mind, we now prove the first main result of our paper.

\begin{theorem}
\label{main theorem}
Given an infinite set of sites $T$ with a $J$, suppose that $\{T_i\}_{i = 1}^{\infty}$ is a pairwise-disjoint collection of subsets of $T$ such that each $T_i$ has the possibility of being the module $M$.  Then, for any $\epsilon >0$ and for any $\delta >0$, there exists a $k_0$ such that for all $k > k_0$, if $\Lambda \subseteq T$ and $\bigcup\limits_{i=1}^{k}T_i \subseteq \Lambda$, then \[P \left [ |D_0(\Lambda,J|_{B(\Lambda)})| > 2^{^{k f(p) p_s^{|M|} p_b^{|B(M)|} (1 - \epsilon)}} \right ] > 1 - \delta.\]
\end{theorem}
 
\begin{proof}
Let $\epsilon >0$ and $\delta >0$ be given.  Since the probability of any $T_i$ being the module $M$ is $f(p) p_s^{|M|} p_b^{|B(M)|}$, the Law of Large Numbers furnishes a $k_0 \in \Nl$ such that for all $k > k_0$, \[P \left [ \frac{n}{k} > f(p) p_s^{|M|} p_b^{|B(M)|}(1 - \epsilon) \right ] > 1 - \delta,\] where $n$ is the number of the first $k$ $T_i$'s that are modules in $T$.  Now let $\Lambda \subseteq T$ be given such that $\bigcup\limits_{i=1}^{k}T_i \subseteq \Lambda$.  Then, there are at least $n$ pairwise disjoint modules $M$ in $\Lambda$, and hence $|D_0(\Lambda,J|_{B(\Lambda)})| \geq 2^n$.  The conclusion now follows.
\end{proof}

\subsection{Existence of Modules}

We next wish to demonstrate the existence of modules for square, triangular, and hexagonal lattices.  Then, using Theorem \ref{main theorem}, we will calculate explicitly a lower bound for each of these lattice types with the restriction that we have a random $J$ configuration---the probability that a bond is a negative $J$ bond is $\frac{1}{2}$---and that the lattice is neither bond nor site diluted.  That is, $p_s = p_b = 1$.  But we will first need to calculate $f(\frac{1}{2})$ for each of the modules.  So as to make the calculation of $f(\frac{1}{2})$ a trivial task, we must introduce the notion of plaquettes and frustrated plaquettes.  If we have a lattice $\Lambda$, a \emph{plaquette} is the basic building block of this lattice.  For example, for a square, triangular, or hexagonal lattice, a plaquette is a set of four, three, or six sites, respectively, that are the vertices of the smallest square, triangle, or hexagon, respectively.  Note that a plaquette can be realized as the range of a closed curve.  Now suppose we have a plaquette in which the product of the $J$'s along this closed curve is $-1$.  Then, by Theorem \ref{basis}, no matter what spin state is put on $\Lambda$, there must be at least one unhappy bond in this plaquette. We therefore call such a plaquette in which the product of the $J$'s is $-1$ a \emph{frustrated plaquette}.  With this terminology and the following lemma, we will be able to compute $f(p)$ for random $J$'s.

\begin{lemma}
Let $s$ be a closed curve of length $n$ with $q < n$ of its $J$ values specified.  Then, the number of ways to get an odd number of negative $J$ bonds along $s$ is equal to the number of ways to get an even number of negative $J$ bonds along $s$.
\end{lemma}

\begin{proof}
Without loss of generality, suppose that of the $q$ bonds already specified, an odd number are negative $J$ bonds.  We can also without loss of generality assume that $n - q$ is even.  Then, the number of ways to get an odd number of negative $J$ bonds is $$ {n - q \choose 0} + {n - q \choose 2} + \cdots + {n - q \choose n - q}.$$  The number of ways to get an even number of negative $J$ bonds is $$ {n - q \choose 1} + {n - q \choose 3} + \cdots + {n - q \choose n - q -1}.$$  However, it is a well known fact that these two are equal.
\end{proof}

Given a random $J$ configuration, this lemma shows us that as long as there is at least one $J$ value not specified on a plaquette, then the probability that this plaquette is frustrated is $\frac{1}{2}$, and the probability that it is unfrustrated is $\frac{1}{2}$.  Thus, given a set of plaquettes whose frustrations are specified on exactly $m$ of these plaquettes, it follows that the probability of this frustrated plaquette configuration occurring is $ \left (\frac{1}{2} \right )^m$, provided of course that we can snake around in such a way that we have at least one degree of freedom for each plaquette.  Therefore, $f(\frac{1}{2}) = \left (\frac{1}{2} \right )^m$.

Having calculated $f(\frac{1}{2})$, we now proceed to prove the existence of modules.  We begin by proving the existence of a module for a square lattice.

\begin{theorem}
Let $\Lambda$ be a square lattice and suppose $M \subseteq \Lambda$ is the set of $25$ sites with any $J_M$ that gives the plaquettes $p_1,\ldots,p_{14}$ the frustrations as indicated in Picture 2.  Then, $M$ is a module in $\Lambda$.
\end{theorem}

\setlength{\unitlength}{1.25cm}
\begin{figure}[ht]
\begin{picture}(7,8)(-1.25,-3)


\put(4.27,4.7){\large Picture 2}
\multiput(0,0)(0,1){5}{\line(1,0){4}}          
\multiput(0,0)(1,0){5}{\line(0,1){4}}

\small

\put(1.05,4.05){1}                             
\put(2.05,4.05){2}
\put(3.05,4.05){3}
\put(4.05,4.05){4}
\put(0.05,3.05){5}
\put(1.05,3.05){6}
\put(2.05,3.05){7}
\put(3.05,3.05){8}
\put(4.05,3.05){9}
\put(0.05,2.05){10}
\put(1.05,2.05){11}
\put(2.05,2.05){12}
\put(3.05,2.05){13}
\put(4.05,2.05){14}
\put(0.05,1.05){15}
\put(1.05,1.05){16}
\put(2.05,1.05){17}
\put(3.05,1.05){18}
\put(4.05,1.05){19}
\put(0.05,0.05){20}
\put(1.05,0.05){21}
\put(2.05,0.05){22}
\put(3.05,0.05){23}

\multiput(0.5,1.5)(1,1){3}{\circle{.5}}        
\multiput(1.5,0.5)(1,1){3}{\circle{.5}}

\normalsize

\put(1.43,3.42){1}                             
\put(2.43,3.42){2}
\put(3.43,3.42){3}
\put(0.43,2.42){4}
\put(1.43,2.42){5}
\put(2.43,2.42){6}
\put(3.43,2.42){7}
\put(0.43,1.42){8}
\put(1.43,1.42){9}
\put(2.34,1.42){10}
\put(3.34,1.42){11}
\put(0.34,0.42){12}
\put(1.34,0.42){13}
\put(2.34,0.42){14}

\put(6.1,0){\begin{picture}(7,0)


\multiput(0,0)(0,1){5}{\line(1,0){4}}          
\multiput(0,0)(1,0){5}{\line(0,1){4}}

\small

\put(1.05,4.05){1}                             
\put(2.05,4.05){2}
\put(3.05,4.05){3}
\put(4.05,4.05){4}
\put(1.05,3.05){6}
\put(2.05,3.05){7}
\put(3.05,3.05){8}
\put(4.05,3.05){9}
\put(1.05,2.05){11}
\put(2.05,2.05){12}
\put(3.05,2.05){13}
\put(4.05,2.05){14}
\put(1.05,1.05){16}
\put(2.05,1.05){17}
\put(3.05,1.05){18}
\put(4.05,1.05){19}

\linethickness{1mm}                    	     

\put(3,3){\line(0,1){1}}                      
\put(4,3){\line(0,1){1}}
\put(1,1){\line(0,1){1}}
\put(2,1){\line(0,1){1}}
\put(1,3){\line(1,0){1}}
\put(1,4){\line(1,0){1}}
\put(3,1){\line(1,0){1}}
\put(3,2){\line(1,0){1}}
\multiput(1.5,1.5)(0,2){2}{\line(1,0){2}}     
\multiput(1.5,1.5)(2,0){2}{\line(0,1){2}}
\multiput(1.5,3.5)(2,0){2}{\circle*{0.08}}    
\multiput(1.5,1.5)(2,0){2}{\circle*{0.08}}
\multiput(2,2)(1,0){2}{\circle*{0.15}}        
\multiput(2,3)(1,0){2}{\circle*{0.15}}

\end{picture}}

\put(0.42,-0.9){\parbox[t]{11.8cm}{In pictures 2-4, numbers at vertices refer to site numbers and numbers in the centers of plaquettes refer to plaquette numbers.  If a plaquette number is circled, then that plaquette is specified as frustrated; otherwise it is specified as unfrustrated.  The frustration of all plaquettes that do not have a number is left unspecified.}}

\end{picture}
\end{figure}

\begin{proof}
Suppose $\sigma_0$ is a ground state for $\Lambda$ with any $J$ such that $J|_{B(M)} = J_M$, and suppose that $M$ is not a module in $\Lambda$.  Then, if $S \subseteq M$ is nonempty, it follows that $S$ is not entropic relative to $U_0 = U(\Lambda,J,\sigma_0)$.

Define $M^{\prime} = \{s_1,\ldots,s_{23}\}$.  By Theorem \ref{entropic set theorem} we know that any site $s_i \in M^{\prime}$ can be involved in at most one unhappy bond in $U_0$; for if $s_i$ were involved in more, then the set $\{s_i\}$ would be entropic relative to $U_0$.  Now consider the plaquette $p_5$.  Since $p_5$ is frustrated, we know that there is at least one bond in $p_5$ that is also in $U_0$.  We thus have four cases:

Case(1) $b_{6,7} \in U_0$

By repeated application of Theorem \ref{basis} to the plaquettes $p_1$, $p_2$, $p_3$, $p_7$, $p_{11}$, $p_{10}$, $p_9$, in that order, we deduce that the following bonds are in $U_0$: $b_{1,2}$, $b_{3,8}$, $b_{4,9}$, $b_{13,14}$, $b_{18,19}$, $b_{12,17}$, $b_{11,16}$, as seen in Picture 2.  We now observe that the set $\{s_7,s_8,s_{12},s_{13}\}$ is entropic relative to $U_0$ which is a contradiction.  Therefore, Case (1) is not possible.

Following the structure of the reasoning in Case (1), we similarly find that none of the cases (2) $b_{6,11} \in U_0$, (3) $b_{11,12} \in U_0$, or (4) $b_{7,12} \in U_0$ is possible.  But this contradicts the fact that at least one of these cases must hold.  Therefore, $M$ is a module in $\Lambda$.
\end{proof}

Before we continue, there are two important things we need to be clear about.  First, thus far we have only shown that \emph{if} there is a $J_M$ that yields the configuration of frustrated plaquettes in Picture 2, then $M$ is a module in $\Lambda$.  To be able to apply Theorem \ref{main theorem}, we need to further show the existence of such a $J_M$.  As with each of our modules, however, the existence of such a $J_M$ is a trivial fact, and we leave its verification to the reader.  Second, to apply Theorem \ref{main theorem} with $p = \frac{1}{2}$, we need to verify that for each of our modules, we can snake around the module in such a way that we have at least one degree of freedom for each plaquette in the module.  This too is an easy exercise that we leave to the reader.

Having said this, we next observe that if we rotate the module in Picture 2 by $90$ degrees, then by symmetry, this will be another module.  We thus have two modules for a square lattice.

We now note that we can define a subset $\Lambda$ of a square lattice that can be partitioned by $5 \times 5$ subsets, and in such a way that $\Lambda$ can have either free, cylindrical, or toroidal boundary conditions.  Thus, since both of our modules have $25$ sites and we have specified the frustration of $14$ plaquettes in each, it follows that the probability that any given $5 \times 5$ subset in the partition is one of our modules is $2kf(\frac{1}{2}) = \frac{|\Lambda|}{25 \cdot 2^{13}}$.  Thus, we can take a number arbitrarily close to $2^{^{\frac{|\Lambda|}{25 \cdot 2^{13}}}}$ as a lower bound for the ground state degeneracy of a square lattice.

We now demonstrate the existence of a module for a triangular lattice.  The module is seen in Picture 3.  The proof that this is a module closely resembles the previous proof.

\begin{theorem}
Let $\Lambda$ be a triangular lattice and suppose $M \subseteq \Lambda$ is the set of $21$ sites with any $J_M$ that gives the plaquettes $p_1,\ldots,p_{19}$ the frustrations as indicated in Picture 3.  Then, $M$ is a module in $\Lambda$.
\end{theorem}

\setlength{\unitlength}{1.92cm}
\begin{figure}[ht]
\begin{picture}(7,5.9)(-1.9,-0.2)

\put(2,5.3){\large Picture 3}


\put(4,0){\line(1,2){0.5}}          
\put(1,0){\line(-1,2){0.5}}
\put(2,4){\line(1,0){1}}
\put(3,0){\line(1,2){1}}
\put(2,0){\line(1,2){1.5}}
\put(1,0){\line(1,2){2}}
\put(0,0){\line(1,2){2.5}}
\put(2,0){\line(-1,2){1}}
\put(3,0){\line(-1,2){1.5}}
\put(4,0){\line(-1,2){2}}
\put(5,0){\line(-1,2){2.5}}
\put(0,0){\line(1,0){5}}
\put(0.5,1){\line(1,0){4}}
\put(1.5,3){\line(1,0){2}}
\put(1,2){\line(1,0){3}}


\put(1.44,3.1){1}                   
\put(2.44,3.1){2}
\put(3.44,3.1){3}
\put(0.94,2.1){4}
\put(1.94,2.1){5}
\put(2.94,2.1){6}
\put(3.94,2.1){7}
\put(1.44,1.1){8}
\put(2.44,1.1){9}
\put(3.39,1.1){10}
\put(1.9,0.1){11}
\put(2.9,0.1){12}

\put(2.5,0.33){\circle{0.5}}        
\put(2.5,1.7){\circle{0.5}}
\put(3.5,2.35){\circle{0.5}}
\put(1.5,2.35){\circle{0.5}}


\put(1.95,3.3){1}                   
\put(2.95,3.3){2}
\put(1.45,2.3){3}
\put(1.95,2.6){4}
\put(2.45,2.3){5}
\put(2.95,2.6){6}
\put(3.45,2.3){7}
\put(0.9,1.3){8}
\put(1.45,1.6){9}
\put(1.9,1.3){10}
\put(2.4,1.6){11}
\put(2.9,1.3){12}
\put(3.4,1.6){13}
\put(3.9,1.3){14}
\put(1.4,0.3){15}
\put(1.9,0.6){16}
\put(2.4,0.3){17}
\put(2.9,0.6){18}
\put(3.4,0.3){19}

\end{picture}
\end{figure}

\begin{proof}
Let $\sigma_0$ be a ground state for $\Lambda$ with any $J$ such that $J|_{B(M)} = J_M$, and suppose that $M$ is not a module in $\Lambda$.  Define $M^{\prime} = \{s_1,\ldots,s_{12}\}$.  Then, any site in $M^{\prime}$ can be involved in at most two unhappy bonds.  Since plaquette $p_{11}$ is frustrated, it has at least one unhappy bond.  By symmetry, we can assume this bond to be $b_{56}$.  We then see that plaquette $p_5$ needs one more unhappy bond since it is unfrustrated.  By symmetry, we can take this bond to be $b_{25}$.  Next, since $p_4$ is unfrustrated and $p_3$ is frustrated, we see that the bonds $b_{12}$ and $b_{14}$ must be unhappy.  We then see that plaquette $p_1$ is unfrustrated.  Since it already has one unhappy bond, it must have one more.  We therefore have an entropic set consisting of either site $s_1$ or $s_2$, which is a contradiction.  Thus, there is indeed a non-trivial entropic set in $M$ relative to $\sigma_0$, and hence $M$ is a module in $\Lambda$.
\end{proof}

We now observe that we can define a subset $\Lambda$ of a triangular lattice that can be partitioned by $M$, and in such a way that $\Lambda$ can have any of the three boundary conditions we are considering.  Thus, since $|M| = 21$ and we have specified the frustration of $19$ plaquettes in $M$, we see that for a random $J$, a lower bound for the ground state degeneracy of a triangular lattice can be taken arbitrarily close to $2^{^{\frac{|\Lambda|}{21 \cdot 2^{19}}}}$.

Let us now turn our attention to hexagonal lattices.

\begin{theorem}
Let $\Lambda$ be a hexagonal lattice and suppose $M \subseteq \Lambda$ is the set of $54$ sites with any $J_M$ that gives the plaquettes $p_1,\ldots,p_{19}$ the frustrations as indicated in Picture 4.  Then, $M$ is a module in $\Lambda$.
\end{theorem}

\setlength{\unitlength}{1cm}
\begin{figure}[ht]
\begin{picture}(10,10.4)(-3.75,-0.4)


\put(3,9.5){\Large Picture 4}

\multiput(2,0.86)(0,1.73){5}{\line(1,0){1}}          
\multiput(5,0.86)(0,1.73){5}{\line(1,0){1}}

\multiput(0.5,1.73)(0,1.73){4}{\line(1,0){1}}
\multiput(3.5,0)(0,1.73){6}{\line(1,0){1}}
\multiput(6.5,1.73)(0,1.73){4}{\line(1,0){1}}

\multiput(2,0.86)(0,1.73){4}{\line(-1,2){0.44}}
\multiput(5,0.86)(0,1.73){5}{\line(-1,2){0.44}}
\multiput(8,2.59)(0,1.73){3}{\line(-1,2){0.44}}

\multiput(0.5,1.73)(0,1.73){3}{\line(-1,2){0.44}}
\multiput(3.5,0)(0,1.73){5}{\line(-1,2){0.44}}
\multiput(6.5,1.73)(0,1.73){4}{\line(-1,2){0.44}}

\multiput(1.5,1.73)(0,1.73){4}{\line(1,2){0.44}}
\multiput(4.5,0)(0,1.73){5}{\line(1,2){0.44}}
\multiput(7.5,1.73)(0,1.73){3}{\line(1,2){0.44}}

\multiput(0,2.59)(0,1.73){3}{\line(1,2){0.44}}     
\multiput(3,0.86)(0,1.73){5}{\line(1,2){0.44}}
\multiput(6,0.86)(0,1.73){4}{\line(1,2){0.44}}

\put(8,2.7){54}                             
\put(8,4.43){26}
\put(8,6.16){28}

\put(7.6,1.63){53}
\put(7.6,3.36){25}
\put(7.6,5.09){27}
\put(7.6,6.82){29}

\put(6.5,1.83){52}
\put(6.5,3.56){24}
\put(6.5,5.29){8}
\put(6.5,7.02){30}

\put(6.1,0.77){51}
\put(6.1,2.5){23}
\put(6.1,4.23){7}
\put(6.1,5.96){9}
\put(6.1,7.69){31}

\put(5,0.97){50}
\put(5,2.7){22}
\put(5,4.43){1}
\put(5,6.16){10}
\put(5,7.89){32}

\put(4.6,-0.09){49}
\put(4.6,1.64){21}
\put(4.6,3.37){6}
\put(4.6,5.1){2}
\put(4.6,6.83){11}
\put(4.6,8.56){33}

\put(3.5,0.11){48}
\put(3.5,1.84){20}
\put(3.5,3.57){5}
\put(3.5,5.3){3}
\put(3.5,7.03){12}
\put(3.5,8.76){34}

\put(3.1,0.77){47}
\put(3.1,2.5){19}
\put(3.1,4.23){4}
\put(3.1,5.96){13}
\put(3.1,7.69){35}

\put(2,0.97){46}
\put(2,2.7){18}
\put(2,4.43){16}
\put(2,6.16){14}
\put(2,7.89){36}

\put(1.6,1.63){45}
\put(1.6,3.36){17}
\put(1.6,5.09){15}
\put(1.6,6.82){37}

\put(0.5,1.83){44}
\put(0.5,3.56){42}
\put(0.5,5.29){40}
\put(0.5,7.02){38}

\put(0.1,2.5){43}
\put(0.1,4.23){41}
\put(0.1,5.96){39}

\multiput(1,2.59)(0,1.73){3}{\circle{.85}}           
\multiput(7,2.59)(0,1.73){3}{\circle{.85}}
\multiput(4,0.86)(0,3.46){3}{\circle{.85}}
\multiput(2.5,1.73)(0,5.19){2}{\circle{.85}}
\multiput(5.5,1.73)(0,5.19){2}{\circle{.85}}

\Large
\put(3.9,4.15){1}                                    
\put(3.75,7.61){11}
\put(3.75,0.69){17}
\put(6.9,4.15){8}
\put(0.75,4.15){14}
\put(6.9,5.88){9}
\put(6.75,2.42){19}
\put(0.75,5.88){13}
\put(0.75,2.42){15}
\put(2.25,1.56){16}
\put(5.25,1.56){18}
\put(2.25,6.75){12}
\put(5.25,6.75){10}
\put(5.4,5.02){2}
\put(5.4,3.29){7}
\put(3.9,5.88){3}
\put(3.9,2.42){6}
\put(2.4,5.02){4}
\put(2.4,3.29){5}

\end{picture}
\end{figure}

\begin{proof}
Let $\sigma_0$ be a ground state for $\Lambda$ with any $J$ such that $J|_{B(M)} = J_M$, and suppose that $M$ is not a module in $\Lambda$.  Letting $U_0 = U(\Lambda,J,\sigma_0)$, we first note that if $S$ is any subset of $M$, then $|B_S \setminus U_0| \geq |B_S \cap U_0|$ since $\sigma_0$ is a ground state.  Also, since each site $s \in M$ is involved in at most three bonds, it follows that at most one of the bonds containing $s$ is in $U_0$.

Now consider plaquette $p_1$.  Since $p_1$ is frustrated, at least one of the bonds in $p_1$ must be in $U_0$.  First consider the case that $b_{23} \in U_0$.

We now observe that if $b_{10,11} \in U_0$, then $S_{10,11} = \{s_2,s_{10}\}$ is entropic relative to $U_0$, while if $b_{12,13} \in U_0$, then $S_{12,13} = \{s_3,s_{13}\}$ is entropic relative to $U_0$.  In view of this, and the fact that $p_3$ is unfrustrated, we conclude that $b_{11,12} \in U_0$.

Using reasoning similar to that just used, together with the fact that $p_{12}$ is frustrated, we see that either $b_{14,37} \in U_0$ or $b_{36,37} \in U_0$.  If $b_{14,37} \in U_0$, then $S_{14,37} = \{s_3,s_{12},s_{13},s_{14}\}$ is entropic relative to $U_0$.  Hence, $b_{36,37} \in U_0$.

From this and the fact that $p_{13}$ is frustrated, we either have $b_{15,40} \in U_0$ or $b_{39,40} \in U_0$.  If $b_{15,40} \in U_0$, then $S_{15,40} = \{s_3,s_{12},s_{13},s_{14},s_{15},s_{37}\}$ is entropic relative to $U_0$.  Thus, $b_{39,40} \in U_0$.

Consequently, since $p_{14}$ is frustrated, either $b_{16,17} \in U_0$ or $b_{17,42} \in U_0$.  If $b_{16,17} \in U_0$, then defining $S_{16,17} = \{s_3,s_4,s_{12},s_{13},s_{14},s_{15},s_{16},s_{37},s_{40}\}$ shows that $|B_{S_{16,17}} \setminus U_0| < |B_{S_{16,17}} \cap U_0|$, which is a contradiction.  Thus, $b_{17,42} \in U_0$.  

However, we now observe that $S_{17,42} = S_{16,17} \cup \{s_{17}\}$ is entropic relative to $U_0$, and so we conclude that $b_{17,42} \notin U_0$, which is a contradiction.

Therefore, $b_{23} \notin U_0$, and hence, by symmetry, none of the bonds in $p_1$ is in $U_0$, contradicting the fact that $p_1$ is frustrated.

Therefore, $M$ is a module in $\Lambda$.
\end{proof}

As with the triangular lattice, we see that we can define a subset $\Lambda$ of a hexagonal lattice that can be partitioned by $M$ and that can have any of the boundary conditions we are considering.  Therefore, since $|M| = 54$ and we have specified the frustration of $19$ plaquettes in $M$, we see that for a random $J$, a lower bound for the ground state degeneracy for a hexagonal lattice can be taken arbitrarily close to $2^{^{\frac{|\Lambda|}{54 \cdot 2^{19}}}}$.

\Section{Conclusion}

We now summarize our results of the lower bounds for the various lattices assuming a random $J$ and no dilution of sites or bonds.  Note that in every case, a lower bound is of the form $2^{^{c |\Lambda| (1 - \epsilon)}}$ for some positive constant $c$.  This constant $c$ is in fact a lower bound for the \emph{ground state entropy density} of the lattice $\Lambda$, that is, the ground state entropy per unit volume, defined by \[\hat{S}_0(\Lambda,J) = \frac{\log_2(|D_0(\Lambda,J)|)}{|\Lambda|}.\]  We therefore have the following lower bounds for the ground state entropy densities of square, triangular, and hexagonal lattices, with a random $J$, and with either free, cylindrical, or toroidal boundary conditions:

\noindent
For a square lattice we have \[c = \frac{1}{25 \cdot 2^{13}} \approx 4.883 \times 10^{-6}.\]  For a triangular lattice we have \[c = \frac{1}{21 \cdot 2^{19}} \approx 9.083 \times 10^{-8}.\]  For a hexagonal lattice we have \[c = \frac{1}{54 \cdot 2^{19}} \approx 3.532 \times 10^{-8}.\]

We should now note that the ground state entropy density $\hat{S}_0$---as calculated via the Ising Model---is not necessarily the same as the ground state entropy density observed in the laboratory.  Indeed, the temperature needed to put a system into its ground state may be unattainable in the laboratory.  Hence, in the laboratory we take the thermodynamic limit first, and then we take the $T \rightarrow 0$ limit.  Let us denote this ``observed'' ground state entropy by $\hat{S}$.  However, the Ising Model is a model at zero temperature.  Thus, $\hat{S}_0$ is obtained by taking the limit $T \rightarrow 0$ and then the thermodynamic limit.  This interchange of limits is cause for concern.  However, Aizenman and Lieb [3] showed that $$\hat{S} = \lim_{\Lambda \uparrow \infty} \sup \hat{S}_0,$$ where the supremum is taken over all boundary conditions.  Therefore, the values of $c$ we obtained for the various lattices are indeed lower bounds for the observed ground state entropies in the laboratory---the physically meaningful quantity.

In closing, we emphasize that the formalism presented here is in no way restricted to the specific lattices or boundary conditions we have considered.  Rather, it is a general method to find a lower bound for the ground state degeneracy of an arbitrary lattice with arbitrary boundary conditions.

Also, while we didn't make explicit reference to the concept of a ``dual lattice'' or minimizing ``path length'' to minimize energy, it is these concepts that began our investigation, offered much insight into the problem, and led to the formalism given here.  We feel that it would be very helpful to become acquainted with these concepts in applying this formalism, and for this purpose we refer the reader to [4] and [5], where these concepts make one of their earliest appearances in the literature.

\section*{Acknowledgements}
We express our gratitude towards P. Contucci and B. Nachtergaele for their help and patience.  This material is based on work supported by the National Science Foundation Research Experiences for Undergraduates under Grant No. DMS 9706599, DMS 0070774, and DMS 9802122.

\section*{References}
[1] Avron J.E., Roepstorff G., and Schulman L.S. 1981 \emph{Journal of Statistical}
\newline
\mbox{ } \mbox{ } \emph{Physics} \bf 26 \normalfont 25-36
\newline
[2] Loebl M. and Vondrak J. \emph{A Theory of Frustrated Degeneracy}, KAM preprint series no. 
\newline
\mbox{ } \mbox{ } 201-517, Department of Applied Mathematics, Charles University, Prague, 2000
\newline
[3] Aizenman M. and Lieb E.H. 1981 \emph{Journal of Statistical Physics} \bf 24 \normalfont 279-297
\newline
[4] Wegner F. 1971 \emph{Journal of Mathematical Physics} \bf 10 \normalfont 2259-2272
\newline
[5] Toulouse G. 1977 \emph{Communications on  Physics} \bf 2 \normalfont 115-119

\end{document}